\documentclass{iopart}

\usepackage{graphicx}
\usepackage[square,sort&compress]{natbib}

\begin{document}

\title[Parameter estimation of spinning binaries using MCMC]{Parameter estimation of spinning binary inspirals using Markov-chain Monte Carlo}

\author{Marc van der Sluys$^{1}$, Vivien Raymond$^{1,2}$, Ilya Mandel$^{1}$, 
  Christian R\"{o}ver$^{3,4}$, Nelson Christensen$^{5}$, Vicky Kalogera$^{1}$, 
  Renate Meyer$^{3}$, Alberto Vecchio$^{1,6}$}
\address{$^1$ Physics \& Astronomy, Northwestern University, Evanston IL, USA}
\address{$^2$ Universit\'{e} Louis Pasteur - Strasbourg I, Strasbourg, France}
\address{$^3$ Statistics, University of Auckland, New Zealand}
\address{$^4$ Max-Planck-Institut f\"ur Gravitationsphysik, Hannover, Germany}
\address{$^5$ Physics \& Astronomy, Carleton College, Northfield MN, USA}
\address{$^6$ Physics \& Astronomy, U. of Birmingham, Edgbaston, Birmingham, UK}
\ead{sluys@northwestern.edu}

\begin{abstract}
  We present a Markov-chain Monte-Carlo (MCMC) technique to study the source
  parameters of gravitational-wave signals from the inspirals of stellar-mass compact
  binaries detected with ground-based gravitational-wave detectors such
  as LIGO and Virgo, for the case where spin is present in 
  the more massive compact object in the binary.
  We discuss aspects of the MCMC algorithm that allow us to sample the 
  parameter space in an efficient way.  
  We show sample runs that illustrate the possibilities of our MCMC code and
  the difficulties that we encounter.
\end{abstract}

\pacs{02.50.-r, 02.70.Uu, 04.30.Tv, 04.80.Nn, 95.85.Sz}
\submitto{Classical and Quantum Gravity}

\section{Introduction}
\label{sec:intro}

Inspirals of stellar-mass compact binaries induced by gravitational radiation are 
among the most promising gravitational-wave sources for ground-based laser 
interferometers, such as LIGO~\cite{1992Sci...256..325A,statusofligo2006}
and Virgo~\cite{statusofvirgo2007}. If such a binary contains a black hole, 
it is believed to be spinning moderately~\cite{2007astro.ph..3131B}.  A 
spinning black hole causes the binary orbit to precess, introducing phase and 
amplitude modulations in the gravitational-wave signal. This should be taken into account in the 
analysis of the signal.  The accuracy with which the binary parameters can be estimated
is of significant astrophysical interest.

We developed a code which implements a Markov-chain Monte-Carlo (MCMC) 
technique~\cite{gilks_etal_1996} to compute the posterior probability-density 
functions (PDFs) of the source parameters.  This code is a modification of an 
earlier parameter-estimation code for analysis on binaries without 
spin~\cite{2007PhRvD..75f2004R, RoeverThesis2007}.  In addition to including 
post-Newtonian gravitational waveforms with a single spinning 
object~\cite{1994PhRvD..49.6274A}, we have also implemented a number of 
improvements designed to make the parameter-space exploration more 
efficient, such as parallel tempering.

This paper is organised as follows.  In section~\ref{sec:implement}, we 
describe the implementation of the MCMC algorithm in the code and related 
details, such as data handling, waveform choice, update proposals, and 
parallel tempering.  Section~\ref{sec:simulation} contains a discussion 
of some sample runs obtained with our MCMC code. In 
section~\ref{sec:conclfuture} we summarise our conclusions, comment on 
existing issues, and describe planned future improvements.

\section{Implementation of MCMC}
\label{sec:implement}

The code we use to estimate the parameters of a binary inspiral
with a spinning member is based on an earlier code that was used by some of
us for the case where no spin is present~\cite{RoeverThesis2007}. 
In this section, we describe some of the features
that were taken from the earlier non-spinning MCMC code, as well as some 
features that were introduced in the present code for use on inspirals 
with a single spinning object.

\subsection{Data handling}
\label{sec:datahandle}

For this study, we inject simulated waveforms with parameters of our 
choice into a stretch of simulated Gaussian, stationary noise at the 
designed sensitivity level for the detectors~\cite{Sigg2004}.  The 
resulting data is windowed, Fourier transformed, and subsequently 
examined by the MCMC analysis.  The details of the data handling, 
Fourier transformation, windowing and PSD estimation can be found 
in~\cite{RoeverThesis2007}.

A stretch of 256 seconds of simulated noise data is used to estimate the 
power-spectral density (PSD) of the noise $S_\mathrm{n}(f)$.  Noise data 
is read in from files in the LIGO/Virgo frame format~\cite{FrameLibrary} 
in the time domain.  Use of the frame-file format also allows us to read 
in real interferometer data and thereby test the analysis on simulated 
hardware-injection signals~\cite{2004CQGra..21S.797B} or analyse any 
candidate events which arise.
In that case, noise estimates would be based on data taken 
close to the time of the signal, without including the 
signal. We have tested our MCMC code on LIGO S5 playground data 
and find that the results are qualitatively similar to the results on 
Gaussian noise.

The signal-to-noise ratio (SNR) of an injected model signal with parameters
$\vec{\lambda}$ as detected by a single detector $i$ is computed as follows:
\begin{equation}
  \rho_i(\vec{\lambda}) = \sqrt{4 \sum_{f=f_\mathrm{low}}^{f_\mathrm{high}} 
    \frac{\left| \tilde{m}(\vec{\lambda},f) \right|^2}{S_\mathrm{n}(f)} \Delta f},
  \label{eq:snr}
\end{equation}
where $\tilde{m}(\vec{\lambda},f)$ is the frequency-domain model waveform, 
$S_\mathrm{n}(f)$ is the noise PSD, the 
sum is computed over the frequency bins between $f_\mathrm{low}$ 
and $f_\mathrm{high}$ and $\Delta f$ is the width of each frequency bin.
The total SNR for a network of $N$ detectors is then given by:

\begin{equation}
  \rho_\mathrm{tot}(\vec{\lambda}) = \sqrt{\sum_{i=1}^N \left(\rho_i(\vec{\lambda})\right)^2}.
  \label{eq:totsnr}
\end{equation}

\subsection{Waveform}

For this phase of our project, we use a simplified waveform
that takes into account post-Newtonian (PN) expansions up to the 1.5\,PN
order in phase and is restricted to the Newtonian order in amplitude.  The 
waveform includes the simple-precession prescription~\cite{1994PhRvD..49.6274A}.
This choice of waveform template allows us to investigate the first-order
effects of spin (spin-orbit coupling), as long as either only one binary
member has spin, or the mass ratio is roughly equal to unity.  In this 
paper, we focus on a fiducial binary consisting of a $10\,M_\odot$ spinning black 
hole and a $1.4\,M_\odot$ non-spinning neutron star.  

The waveform for an inspiral with one spinning object is described by twelve
parameters.  The parameters are: the chirp mass $M_\mathrm{c}$,
symmetric mass ratio $\eta$, spin magnitude $a_\mathrm{spin} \equiv S/M^2$,
the constant angle between spin and orbital angular momentum $\theta_\mathrm{SL}$,
the luminosity distance $d_\mathrm{L}$ and sky position R.A., Dec.,
the time, orbital phase and precession phase at coalescence $t_\mathrm{c}$, 
$\phi_\mathrm{c}$, $\alpha_\mathrm{c}$, and two angles that define the
direction of the total angular-momentum $\vec{J}_0$ of the binary: 
$\theta_\mathrm{J_0}$ and $\phi_\mathrm{J_0}$.

Each waveform template is computed in the time domain, and then windowed and Fourier 
transformed.  The calculation of the likelihood, which measures how well a model 
waveform matches the data, is carried out in the frequency domain.

In this initial study, we focus on waveforms for our fiducial binary
($M_\mathrm{c} \approx 3.0 M_\odot, \eta \approx 0.11$) at a distance of 
$d_\mathrm{L}=13.0\,\mathrm{Mpc}$.  We analyse the effect of spin in an initial 
parameter study~\cite{vds_inpreparation}.
The sky position and orientation of the binary are fixed in this study.

\subsection{Computation of the likelihood}
\label{sec:likelihood}

We follow a Bayesian approach to infer the posterior probability-density
functions (PDFs) of the twelve parameters that describe our waveform.
The PDF of a parameter vector $\vec{\lambda}$ given an observed data set
$d$ follows from Bayes' theorem:
\begin{equation}
  p(\vec{\lambda}|d) = \frac{p(\vec{\lambda})\,p(d|\vec{\lambda})}{p(d)} ~ 
  \propto ~ p(\vec{\lambda})\,L(d|\vec{\lambda}),
  \label{eq:bayes}
\end{equation}
where $p(\vec{\lambda})$ is the {\it prior} distribution of the parameters, 
and $L(d|\vec{\lambda})$ is the {\it likelihood} function.
We calculate the likelihood for a model waveform $\tilde{m}(\vec{\lambda},f)$
with parameters $\vec{\lambda}$ and data set $\tilde{d}(f)$ as measured by 
a detector $i$ in the usual way:
\begin{equation}
  L_i(d|\vec{\lambda})  \propto \exp \left( -2 \int_0^\infty  \frac{\left|\tilde{d}(f) 
    - \tilde{m}(\vec{\lambda},f) \right|^2}{S_\mathrm{n}(f)}\, df  \right).
  \label{eq:likelihood}
\end{equation}
The tildes indicate that both $d$ and $m$ are expressed in the frequency domain.
Since we will be considering the {\it ratio} of likelihoods, we do not
need to take into account the normalisation factor, and it is sufficient to
compute the proportionality in Eq.\,\ref{eq:likelihood}.

Assuming that the noise of different interferometers is independent, the 
expression of the PDF given data from a coherent network of $N$ interferometers 
generalises to:
\begin{equation}
  p(\vec{\lambda}|d) \propto p(\vec{\lambda}) \, \prod_{i=1}^{N} L_i(d|\vec{\lambda}).
  \label{eq:netlikelihood}  
\end{equation}

\subsection{Prior distribution}

We use a prior distribution that is uniform in $\log(d_\mathrm{L})$, 
$\cos(\theta_\mathrm{SL})$, $\sin(\mathrm{Dec})$, 
$\sin(\theta_\mathrm{J_0})$, (the sine is used for parameters defined in 
the domain $[-\frac{\pi}{2},\frac{\pi}{2}]$, the cosine for $\theta_\mathrm{SL}\in [0,\pi]$)
and in the original scales of the remaining parameters.
The allowed ranges for these parameters are between 1 and 6$M_\odot$
for $M_\mathrm{c}$, between 0 and 0.25 for $\eta$, in the range $t_c \pm 
50\mathrm{ms}$, below 100Mpc for $d_\mathrm{L}$, between 0 and 1 for
$a_\mathrm{spin}$, between -1 and 1 for the angles of which we use the
sine or cosine as MCMC parameter, and between 0 and $2\pi$ for all 
other angles.

\subsection{Proposals}

The Markov chain is created as follows. If in the current iteration $i$,
the chain has the location in parameter space (set of waveform parameters, 
or state) $\vec{\lambda}_i$, we propose a random jump $\Delta\vec{\lambda}_i$ 
to the new location $\vec{\lambda}_{i+1} = \vec{\lambda}_i + \Delta\vec{\lambda}_i$.
Since the jump proposal is random, the next state of the chain should 
depend {\it only} on the current state, thus giving the chain its Markovian
property.

We then compute the likelihood for the new state as given by 
Eq.\,\ref{eq:netlikelihood} and determine whether to accept it
by comparing the acceptance probability (the left-hand side in Eq.\,\ref{eq:accept})
to a random number $r$ drawn from a uniform distribution between 0 and 1:
\begin{equation}
  \frac{p(\vec{\lambda}_{i+1})}{p(\vec{\lambda}_{i})} \frac{L(d|\vec{\lambda}_{i+1})}{L(d|\vec{\lambda}_{i})}
  > r.
  \label{eq:accept}
\end{equation}
The jump to state $\vec{\lambda}_{i+1}$ is accepted if Eq.\,\ref{eq:accept} 
is fulfilled.  Otherwise the jump is rejected, the chain keeps the old 
parameter set $\vec{\lambda}_{i+1} = \vec{\lambda}_{i}$ and a new iteration
is started by drawing a new random jump proposal $\Delta\vec{\lambda}_{i+1}$
to a different state $\vec{\lambda}_{i+2}$.  Equation~\ref{eq:accept} 
shows that a new state is always accepted when it improves the product of 
the prior and the likelihood, and that a larger decrease in this product 
means a smaller probability of acceptance.

We use an {\it adaptive} scheme for the proposed jump size~\cite{atchade_rosenthal_2005}.  
The size of the jump proposal for the parameter $\lambda^j$ (the $j$-th 
element of the vector $\vec{\lambda}$) is drawn from a Gaussian distribution 
with width $\sigma_\mathrm{jump}^j$.  Thus, these widths form a vector 
$\vec{\sigma}_\mathrm{jump}$ with the same number of elements as 
$\vec{\lambda}$. The adaptation of the jump size consists of 
increasing $\sigma_\mathrm{jump}^j$ when a jump proposal in the parameter 
$\lambda^j$ is accepted and decreasing it when a proposal is rejected.  
In a typical run, the increase of $\sigma_\mathrm{jump}^j$ is a factor 
of $\sim$8 and the decrease a factor of $\sim$2, which results in the 
target acceptance ratio of about 25\%.

\subsubsection{Uncorrelated proposals}

The default method for choosing a jump proposal is to draw the jump size 
independently in the different dimensions of the parameter space.  This
implies that adaptation is done per parameter as well.  We make these updates
in two categories. The first category contains per-parameter updates, where 
the likelihood is calculated after proposing a jump in one parameter only, 
thus deciding whether to accept the jump for each parameter separately.
The second category involves proposing a jump in all parameters at once
before calculating the likelihood only once.  This is typically done in
10\% of the uncorrelated update proposals.  For both categories of 
uncorrelated update proposals the same vector $\vec{\sigma}_\mathrm{jump}$ 
is used.

\subsubsection{Correlated proposals}
\label{sec:corr}

There can exist strong correlations between parameters, in which case
uncorrelated updates can be very inefficient.  We implemented a
method to calculate the correlations between the parameters of a block of
$n_\mathrm{corr}$ iterations (typically $n_\mathrm{corr}\approx10^4$).
We then draw the subsequent $n_\mathrm{corr}$ jump proposals from 
a multivariate normal distribution that is given by the Cholesky
decomposition of this covariance matrix.

We recompute the covariance matrix and its Cholesky decomposition 
at the end of each block of $n_\mathrm{corr}$ iterations, and decide whether to use 
the new matrix or not by checking how the diagonal elements of the matrix 
have changed.  We find that if we accept each proposed matrix update 
(for which the covariance matrix is positive definite), our proposed
jump sizes may become very small.  This problem does not arise if we 
accept the new matrix only when $\sim 50\%$ of the diagonal elements
have {\it decreased} in value.

The correlated update proposals are always block updates of all twelve 
parameters at once, hence there is a separate $\sigma_\mathrm{jump,\, 
corr}$ for these updates.  In a typical MCMC run, 70--90\% of the update 
proposals are done in a correlated way, and 10--30\% in an 
uncorrelated way.

\subsection{Parallel tempering}
\label{sec:partemp}

The problem that arises when using MCMC for parameter estimation, and 
especially to find the (unknown) modes of the PDFs, is that the chains should 
typically be broad enough to sample the whole allowed parameter ranges, 
while also being able to probe the region of maximum likelihood in a 
detailed way. These two demands are almost mutually exclusive, but
the technique known as {\it parallel tempering} offers a solution.

Parallel tempering consists of several parallel chains that each have a
different {\it `temperature'}.  In addition to the default Markov chain 
with $T=1$, parallel chains of higher temperature are computed.  Hotter 
chains are more likely to accept a jump that decreases the likelihood, 
by adjusting Eq.\,\ref{eq:accept} to accept a jump when
\begin{equation}
  \left( \frac{p(\vec{\lambda}_{i+1})}{p(\vec{\lambda}_{i})} \frac{L(d|\vec{\lambda}_{i+1})}{L(d|\vec{\lambda}_{i})} \right)^{\frac{1}{T}} > r,
  \label{eq:accept_pt}
\end{equation}
where $T\geq1$ is the temperature of the chain.  (Eq.\,\ref{eq:accept_pt} can be viewed 
as the definition of ``temperature''.)  The property of more frequently accepting jumps that
lower the likelihood allows a hot chain to move around in parameter space 
more widely than a cooler chain, thus allowing it to discover different
modes.  Hence, a combination of hot and cool chains 
can probe both wide parameter ranges and the narrow region(s) of maximum 
likelihood.  In order to do so, the chains must be able to exchange information.  
This is done by swapping the parameter sets between two parallel chains with 
$T_m < T_n$ whenever:
\begin{equation}
  \left(\frac{L_n}{L_m}\right)^{\frac{1}{T_m}-\frac{1}{T_n}} > r.
  \label{eq:swapchains}
\end{equation}
Since the likelihood that is needed to determine whether to swap the parameter
sets was already calculated, this decision comes almost for free, and we make 
it for every pair of chains at every iteration.  Output of parallel chains 
with different temperatures for a sample run is shown in 
Fig.\,\ref{fig:parallel_tempering} in Sect.\,\ref{sec:partemp_example}.

\subsubsection{Setting up a temperature ladder}
The temperature ladder is determined by setting the lower temperature to
$T=1$.  This is the only chain that is saved and used to create the PDFs.
One also has to choose a maximum temperature $T_\mathrm{max}$, which is 
typically the lowest temperature that allows the chain to scatter over 
the whole allowed parameter ranges quickly.  In our test runs, we find 
that we need to increase the value of $T_\mathrm{max}$ when injecting a
signal with a higher SNR.  The last quantity to choose
is the number of parallel chains $N_\mathrm{ch}$ in the temperature ladder.
This will be a compromise between high computation speed (low $N_\mathrm{ch}$)
and high swap efficiency for the chains by having small differences between
adjacent temperatures (high $N_\mathrm{ch}$).  The temperatures are then
chosen equidistantly in $\log(T)$.  Our typical setup is $N_\mathrm{ch}\approx 7$ 
and $T_\mathrm{max}\approx30-50$ for SNRs between 10 and 20.

\subsubsection{Sinusoidal temperatures}
The obvious drawback of parallel tempering is that one has to calculate a 
handful of chains, instead of just one.  In order to reduce the number 
of chains in the temperature ladder, we have tested our simulations with 
{\it sinusoidal} temperatures, for all chains with $T>1$.  In order to do 
this, we set up our temperature ladder as before, but now sinusoidally 
oscillate the temperature of each chain $m \neq 1$ with an
amplitude $\Delta T_m$.  We find that the swapping is
efficient when we choose $\Delta T_m = T_m - T_{m-1}$ for each
chain $m>1$, so that the minimum temperature of each chain is equal to
the mean temperature of the next cooler chain.  Furthermore, we make
sure that adjacent chains are in antiphase, so that there is an optimal
overlap at the extrema.  In this setup, we can use 
$N_\mathrm{ch}\approx 4-5$ and $T_\mathrm{max}\approx15-30$ for SNRs between
10 and 20, thus reducing
the computational cost of the MCMC runs with parallel tempering.
We suggest that the period of the temperature variation should not be too
close to $n_\mathrm{corr}$ (see Sect.\,\ref{sec:corr}) and that a too 
short period may endanger the Markovian nature of the chain.  Hence 
we chose a period that is $\sim 5 \times n_\mathrm{corr}$.

\section{MCMC simulations}
\label{sec:simulation}

In a typical MCMC run, we simulate a data set by injecting a 
signal into detector noise, as described in Sect.\,\ref{sec:datahandle}.
This data can be either Gaussian,
stationary noise that is simulated at the designed sensitivity of the
detectors, or real LIGO or Virgo data.  The test simulations
described in this study were all done using synthetic noise.  

For our simulations, we inject the signal with a given set of parameters
into the noise, appropriately projecting it onto one
or more of the following gravitational-wave detectors:
the 4-km LIGO detectors at Hanford, WA (H1), Livingston, LA (L1) in the
USA, or the 3-km Virgo detector near Pisa, Italy.  This way, we can 
do coherent parameter estimation with such a network of detectors,
as described in Sect.\,\ref{sec:likelihood}.

Test simulations that we carry out at the moment focus on a fiducial
binary that consists of a $10\,M_\odot$ spinning black hole and a 
$1.4\,M_\odot$ non-spinning neutron star at a distance of typically
15-20\,Mpc.  
The purpose of these test simulations is twofold.  Firstly, we want 
to establish to which accuracy the source parameters can be measured.
Secondly, our code must be able to find these source parameters when
started from arbitrary values.  These are two different goals that
require slightly different simulations to test in an efficient way.

\subsection{Accuracy of parameter estimation} 
For a fixed network SNR (see Eq.\,\ref{eq:totsnr}), the accuracy with 
which the parameters can be determined depends mainly on three factors: 
the number of detectors in the network, the sky position and orientation 
of the binary (these quantities determine how the network SNR is 
distributed over the different interferometers) and the magnitude and 
the direction of the spin of the black hole. 
We are therefore carrying out a systematic study in which we vary these 
parameters in order to map their influence on the accuracy of the 
parameter estimation.  Because of the large number of parameters that 
is varied, and the timescale of 1-2 weeks that is needed for the chains 
to accumulate a sufficient number of iterations, this is a lengthy process. 
To speed up these simulations, we usually start the Markov chains from 
the true source parameter values which were used for the software injection.  
We must be careful, however: in addition to parameter-estimation uncertainties 
caused by noise, which are quickly measured by seeding the chains with 
the true parameters, there can also be uncertainties due to degeneracies 
or near-degeneracies in the parameter space, which may not become apparent 
in short runs starting from the true parameter values.  
We typically start five serial chains for the same analysis.  Each of these 
Markov chains uses parallel tempering, as described in Sect.\,\ref{sec:partemp}, 
and hence consists of several {\it parallel} chains each with a different 
temperature. Although these serial chains start from the same (true) values, 
they use a different seed for the random-number generator and therefore 
produce different and independent Markov chains.

\subsection{Finding the true parameters}
\begin{figure*}
  \begin{center}
    \resizebox{\textwidth}{!}{
      \includegraphics[angle=-90]{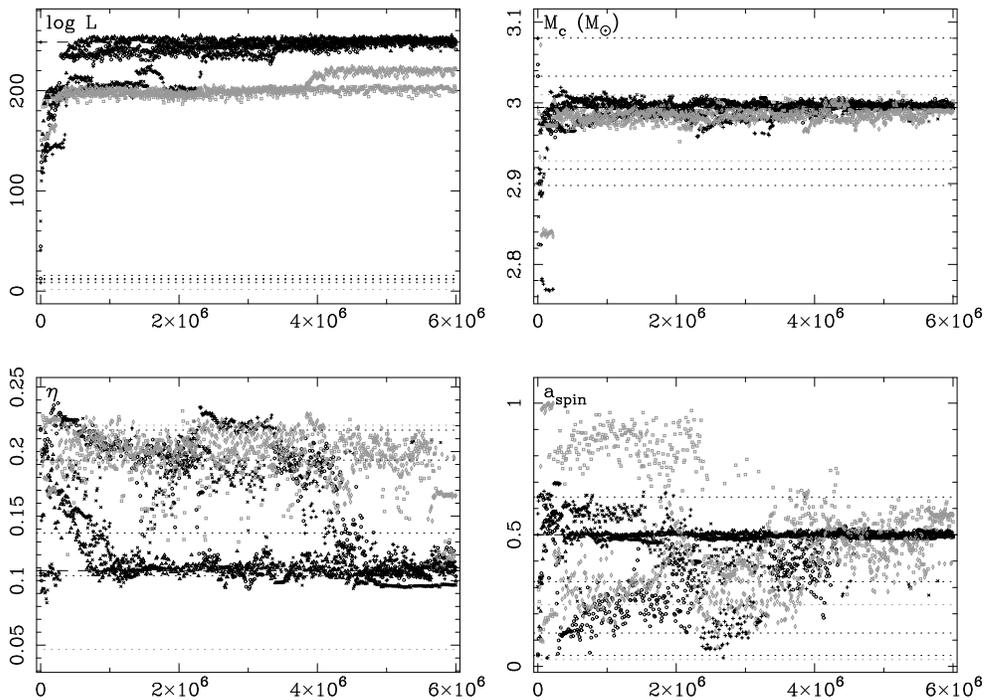}
    }
    \caption{
      Likelihood and Markov chains for the parameter estimation on a
      simulated injection.  The MCMC run consists of six serial chains 
      plotted in two shades of grey.  All of the four black chains 
      find the parameters of the injected signal after $\sim 4.5\times10^6$
      iterations, while the two grey chains have found other, local 
      maxima. The figure shows the difference between the 
	logarithm of the likelihood $\log{L}$ 
      for a waveform with the current parameter values and 
      $\log{L}$ for the null waveform
      (upper-left panel), and the chain projections 
      for the chirp mass $M_\mathrm{c}$ 
      (upper-right), symmetric mass ratio $\eta$ (lower-left) and spin 
      magnitude $a_\mathrm{spin}$ (lower-right). All horizontal axes show
      the iteration number in the chains. The six dotted lines in each 
      panel indicate the starting value of each chain of the corresponding 
      colour.  The dashed black lines are the parameter values of
      the injected signal, and the corresponding value for the likelihood.  
      One out of $\sim 10^4$ calculated iterations is plotted.
    }
    \label{fig:good_bad_chains}
  \end{center}
\end{figure*}

The second purpose of the simulations with our MCMC code is to find the
true source parameters in the case where these are unknown.  In order
to test the ability of our MCMC code to do so, we carry out a semi-blind
analysis, in which the chains are started from offset ({\it i.e.}, non-true)
parameter values.  For the chirp mass and the time of coalescence, these
starting values are drawn from a Gaussian distribution that is centred
on the value of the injected parameter, with a standard deviation of
about $0.1\,M_\odot$ and 30\,ms respectively.  The other ten parameters
are drawn randomly from the allowed ranges (see {\it e.g.}
Fig.\,\ref{fig:good_bad_chains}).  By selecting the starting values for
the chains this way, we model the information that will be available after
a detection trigger is analysed by the LIGO-Virgo data-analysis pipeline.

When starting chains from offset values, we typically use up to ten serial
chains. The reason for using a larger number of chains is that chains may get
stuck at a local maximum in likelihood.  In the case of our semi-blind 
analysis, it is easy to recognise such chains, since we can see whether 
the chains have found the likelihood of the injected signal.  In the case
of a real analysis, we need to be sure that the chains have found the highest
likelihood present.  One way to do this is by starting multiple serial
chains from different positions in parameter space and requiring that they
find the same highest value for the likelihood.  Figure~\ref{fig:good_bad_chains}
shows a run where six chains were started from offset values.  After about
$4.5\times10^6$ iterations, the four black chains have all found the same 
likelihood and parameter values, whereas the two grey chains are exploring
different parts of parameter space.  The latter two chains are clearly at a 
lower value of the likelihood and have therefore found {\it local} maxima in
parameter space.  If the run were continued, the last two chains should
eventually find the parameter values of the injected signal.

\subsection{Example of parallel tempering}
\label{sec:partemp_example}

\begin{figure*}
  \begin{center}
    \resizebox{\textwidth}{!}{
      \includegraphics[angle=-90]{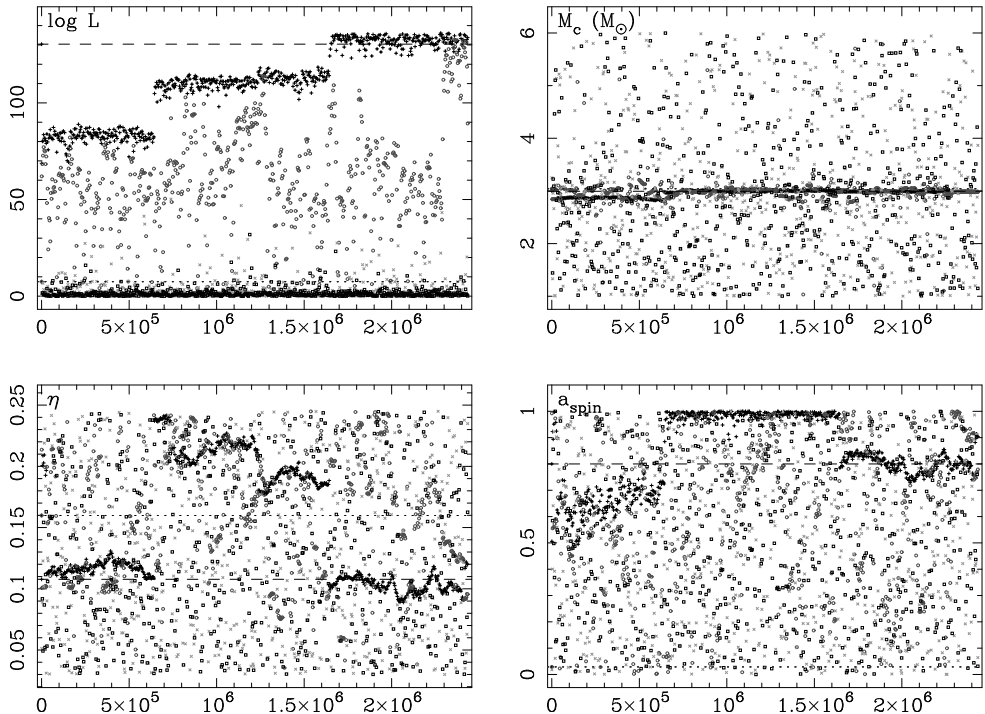}
    }
    \caption{
      Likelihood and Markov chains for chains of different temperature in
      a sample run with parallel tempering.  The panels contain
      the same information as those in Fig.\,\ref{fig:good_bad_chains},
      and the dashed and dotted lines have the same meaning as in that
      figure.
      The chains with $T>1$ have sinusoidal temperatures and chains with 
      increasing temperature are displayed by black
      plusses ($T=1.0$), dark-grey circles ($T_\mathrm{mean}\!\approx\!2.5$), 
      light-grey crosses ($T_\mathrm{mean}\!\approx\!6.3$) and black squares 
      ($T_\mathrm{mean}\!\approx\!16.0$).
      One out of $\sim 4000$ calculated iterations is plotted.
    }
    \label{fig:parallel_tempering}
  \end{center}
\end{figure*}

Figure~\ref{fig:parallel_tempering} shows how the parallel chains with
different temperatures cooperate in the case of parallel tempering. The
figure shows the likelihood and Markov chains for four different
parallel chains.  The chains with $T>1$ have sinusoidal temperatures 
in the ranges 1.0--4.0, 2.5--10.1 and 6.3--25.5.  
The coolest chain (black plusses) has the highest likelihood and
the smallest spread in the parameters, whereas the hottest chain
(black squares) has the lowest likelihood and largest spread.
The spread in likelihood in 
especially the second coolest chain (dark-grey circles) can be attributed
in part to the sinusoidal variation in temperature; whenever the temperature drops,
the chain may climb a nearby `hill' in likelihood, find a high value and
communicate the location of that hill to the cooler chain. The two `sudden' jumps 
for the coolest chain, around iterations $5.2\times10^5$ and $1.52\times10^6$, 
may be explained that way.  In the first jump, the chain moves to the
correct value for the chirp mass, in the second jump, the `true' values
for the mass ratio and spin magnitude are found.  In most MCMC runs, we
start 5--10 of these sets of parallel chains, but only save the output for 
the $T=1$ chains.

\section{Conclusions and future work}
\label{sec:conclfuture} 

We have developed a Markov-chain Monte-Carlo (MCMC) algorithm that we 
use to estimate the twelve physical parameters of the gravitational waves 
emitted during a spinning compact-binary inspiral that can be detected with 
ground-based gravitational-wave observatories like LIGO and Virgo.  In 
Section~\ref{sec:implement}, we discuss many of the implemented features 
that are needed to run this code efficiently.  In 
Section~\ref{sec:simulation} we show examples of MCMC simulations 
carried out with our code.

We are constantly working on improving the efficiency with which the 
Markov chains explore the parameter space.  In particular, a more 
efficient sampler speeds up the search for the true modes of the PDFs
when the Markov chains are started from offset ({\it i.e.} non-true) 
initial parameter values, as they would be in the case of a real 
detection.

When the structure of the likelihood function in the parameter space is 
very rich, and the SNR is high, there are many sharply peaked local 
maxima of the likelihood.  In this case, the MCMC algorithm is likely to 
get stuck on some of these local maxima, as random jumps are unlikely to 
lead from one local maximum to another.  Coherent 
changes in parameters based on an improved analytical understanding of 
the waveform may allow us to efficiently traverse the peaks of this complicated 
parameter space.  We are currently working on gaining a sufficient 
understanding of the harmonic structure of the waveform, which would 
allow us to implement such coherent jumps.

A part of our effort is directed at understanding degeneracies that exist
between parameters, especially the sky position and orientation of the
binary.  For example, we find cases where the PDF for the sky position
is more-or-less uniform over the whole sky, cases where there are multiple
regions in the sky where the binary could be, and cases where one unique
sky position is resolved.  We need to understand better how these degeneracies 
depend on the number of detectors in the network and on the exact configuration 
of that network with respect to the binary position and orientation, and on
the spin of the black hole.

Another part of our work focuses on the implementation of a more realistic, 
higher-order post-Newtonian (PN) waveform, that includes
the spin of both binary components~\cite{2003PhRvD..67j4025B}.  The 
inclusion of the second spin will allow us to
investigate inspirals where both spins are equally important, such as
for double-neutron-star and double-black-hole binaries.  The
implementation of a higher-order PN waveform is expected to increase the
accuracy of parameter estimation and to reduce the bias that inevitably
arises when using approximate waveforms.

At the moment, we are testing our MCMC code regularly on software injections 
into real interferometer data ({\it e.g.} LIGO S5 playground data) and we 
plan to explore the possibility of doing follow-up on candidate events that
come out of the LIGO detection pipeline~\cite{S3S4insprialsearch, S3spinningsearch}.  
We have detailed plans to include this MCMC 
code as a final stage in the LIGO pipeline, in order to provide a post-processing
tool that can be used after a detection.

\section*{Acknowledgements}

This work is partially supported by a Packard Foundation Fellowship, a NASA BEFS grant  
(NNG06GH87G), and a NSF Gravitational Physics grant (PHY-0653321) to VK; NSF Gravitational  
Physics grant PHY-0553422 to NC; Royal Society of New Zealand Marsden Fund grant UOA-204  
to RM and CR; UK Science and Technology Facilities Council grant to AV.  Computations were  
performed on the Fugu computer cluster funded by NSF MRI grant PHY-0619274 to VK.

\bibliographystyle{iopart-num}
\bibliography{gwdaw12}

\end{document}